\documentclass[12pt]{article}
\usepackage[table]{xcolor}
\RequirePackage[OT1]{fontenc}
\RequirePackage{amsthm,amsmath}
\RequirePackage{natbib}
\RequirePackage{graphicx,bm}
\RequirePackage[colorlinks,citecolor=blue,urlcolor=blue]{hyperref}
\usepackage{subcaption}
\usepackage{array}
\usepackage{multirow}
\usepackage{longtable}
\usepackage[tableposition=below]{caption}
\usepackage{booktabs}
\usepackage{ltxtable} 
\usepackage{caption} 
\usepackage{multido}
\usepackage{dcolumn}
\usepackage{lscape}
\definecolor{LRed}{rgb}{1,.8,.8}
\definecolor{MRed}{rgb}{1,.6,.6}
\definecolor{HRed}{rgb}{1,.2,.2}
\usepackage{adjustbox}
\usepackage{tabularx}

\usepackage{rotating}

\usepackage{graphicx, graphics, setspace}
\usepackage{subcaption}
\usepackage{amssymb, url, natbib}
\RequirePackage{amsthm,amsmath}
\RequirePackage{natbib}
\RequirePackage{graphicx,bm}
\RequirePackage[colorlinks,citecolor=blue,urlcolor=blue]{hyperref}
\usepackage{subcaption}
\usepackage{tikz}
\usetikzlibrary{matrix}
\usetikzlibrary{arrows,fit,positioning}
\usepackage{lmodern} 
\usepackage{booktabs}

\usepackage{mathtools}
\usetikzlibrary{fit}
\tikzset{%
  highlight/.style={rectangle,rounded corners,fill=red!15,draw,fill opacity=0.5,thick,inner sep=0pt}
}

\usepackage{xparse}

\NewDocumentCommand{\ceil}{s O{} m}{%
  \IfBooleanTF{#1} 
    {\left\lceil#3\right\rceil} 
    {#2\lceil#3#2\rceil} 
}
\usepackage[letterpaper,right=1in,left=1in,top=1in,bottom=1in]{geometry}
\definecolor{LightCyan}{rgb}{0.88,1,1}

\newcommand{\B}{\mathcal{B}}
\newcommand{\Y}{\mathbf{Y}}

\newcommand{\M}{\mathbf{M}}
\newcommand{\U}{\mathbf{U}}
\newcommand{\V}{\mathbf{V}}

\renewcommand{\u}{\mathbf{u}}
\renewcommand{\v}{\mathbf{v}}

\newcommand{\X}{\mathbf{X}}
\newcommand{\W}{\mathbf{W}}

\renewcommand{\S}{\mathbf{S}}
\renewcommand{\P}{\mathbf{P}}

\newcommand{\normdist}[2]{\ensuremath{\mathcal{N}(#1,#2)}}

\newcommand{\betadist}[2]{\ensuremath{\mathcal{B}eta(#1,#2)}}

\tikzset{
  latentnode/.style  ={draw,minimum width=2.5em, shape=circle,thick, black,fill=white},
  visiblenode/.style ={draw, minimum width=2.5em, shape=circle,thick, black,fill=black!20},
  plate/.style={draw,
    shape=rectangle,
    thick,
    minimum width=4em,
    minimum height=4em,
    align=right,
    inner sep=5em,
    inner ysep=5em,
    label={[xshift=&16pt,yshift=11pt]south east:#1}},
  line/.style={draw, &latex'},
  jump left/.style={draw=none, inner xsep=0pt},
  jump line/.style={line, shorten <=5pt}
}

\usepackage{csquotes}

\begin{document}

\title{Twists and Turns in the US-North Korea Dialogue:
\newline \Large Key Figure Dynamic Network Analysis using News Articles}

\author{Sooahn Shin,
Hyein Yang, and 
Jong Hee Park \\
Seoul National University}

\maketitle
\noindent \textbf{Abstract}\\ 
In this paper, we present a method for analyzing a dynamic network of \emph{key figures} in the U.S.-North Korea relations during the first two quarters of 2018. Our method constructs key figure networks from U.S. news articles on North Korean issues by taking co-occurrence of people's names in an article as a domain-relevant social link. We call a group of people that co-occur repeatedly in the same domain (news articles on North Korean issues in our case) ``key figures'' and their social networks ``key figure networks.'' We analyze block-structure changes of key figure networks in the U.S.-North Korea relations using a Bayesian hidden Markov multilinear tensor model. The results of our analysis show that block structure changes in the key figure network in the U.S.-North Korea relations predict important game-changing moments in the U.S.-North Korea relations in the first two quarters of 2018. 

\vspace*{.3in}


\newpage
\begin{spacing}{2}
\section{Motivation}
News articles contain factual and interpretative information about political events and hence are a highly valuable source of textual data for the study of political science \citep{Druckman2005, Gerber2009, Kim2018, Moeller2014, Moy2004, Gentzkow2010}.  There are two additional aspects of news articles that need to be stressed in political science research. The first is its \emph{sequential} nature. Articles are generated sequentially, closely following everyday political events. Thus, the rise and fall of political issues or turning points in political events can be tracked and predicted by analyzing news articles as \emph{time series text data}. The second, a less well-recognized than the first, aspect of news articles is that they contain valuable information about \emph{social networks} among a wide range of actors such as academics, non-academic experts, lobbyists, think-tanks, private individuals, and politicians. Thus, news articles on a specific issue provide a rich source of dynamic network analysis on political actors involved in the public discourse or policy-decision on the issue. 


The goal of this paper is to present a simple hybrid method that takes advantage of these two aspects of news articles to understand dynamics of political events. The case in our attention is the U.S.-North Korea relations in the first two quarters of 2018. During the entire period of 2017, North Korea had threatened the U.S. that North Korea could launch a nuclear attack on ``the heart of the United States'' if the U.S. attempts a regime change in North Korea. The aggressive tone in the often hostile exchanges further escalated in early 2018. In the New Year address of 2018, Kim Jung Un mentioned a nuclear button under his desk and President Trump replied by mentioning a bigger nuclear button under his finger. After the Winter Olympics, however, both sides withdrew provocative remarks and in March, President Trump accepted an invitation to the U.S.-North Korea summit, which was delivered by a South Korean envoy. Nevertheless in May, President Trump announced the cancellation of the summit, which he retracted only a week later. In June, the two leaders finally met in Singapore and had the first summit since the Korean War.


In this paper, we aim to understand dynamics of the U.S.-North Korea relations by analyzing \emph{dynamics of latent social networks among key figures} in the U.S.-North Korea relations. Our method proceeds in the following order. First, we collect news articles containing the word ``Korea'' from the official sites of 4 major U.S.-based news providers --- \emph{New York Times, Washington Post, Wall Street Journal}, and  \emph{Fox News} --- and build an initial corpus.\footnote{We chose these four news providers for two reasons. First, \emph{Wall Street Journal} , \emph{New York Times}, and \emph{Washington Post} are chosen because they are leading daily newspapers in the U.S., holding a strong international reputation in covering a wide range of global issues. Their daily circulations are 1,180,460, 597,955, and 313.156, respectively as of 2017 (Pew Research Center, ``Newspapers Fact Sheet'' \url{http://www.journalism.org/fact-sheet/newspapers/}). Second,  we added a television news channel, \emph{Fox News}, to this list because of its special status under the Trump administration. More specifically, President Trump repeatedly endorsed \emph{Fox News} for their ``fair" coverage of his presidency (Trump, Donald (@realDonaldTrump), 12:33 PM - Nov 1, 2018) \citep{Gaffey2017}.} Second, we extract from this initial corpus co-occurrence information of $N$ key figures in news articles and construct $N \times N \times T$ co-occurrence tensor (i.e.\ a $T$ array of $N$ by $N$ networks) of key figures.  Last, we apply a Bayesian hidden Markov multilinear tensor model \citep{Sohn2017, Park2017} to the constructed co-occurrence tensor to identify structural changes in key figure networks. 

This hybrid approach can be applied to any type of sequential text data containing proper network-specific (i.e.\ node and link) information and combined with other types of network methods such as community detection or latent space modeling.  We call this hybrid approach Key Figure Dynamic Network Analysis (KF-DNA). The results of our analysis show that block structure changes in the key figure network in the U.S.-North Korea relations predict important turning points in the U.S.-North Korea relations during the first two quarters of 2018. 



\section{Key Figure Dynamic Network Analysis Method}
Our proposed method (KF-DNA) can be summarized in the following three steps: 
\begin{enumerate}
\item Construction of an initial corpus on U.S.-North Korea relations from news articles of four major US news agencies
\item Construction of $N \times N \times T$ co-occurrence tensor (i.e.\ a $T$ array of $N$ by $N$ networks) of key figures in news articles  
\item Bayesian hidden Markov multilinear tensor analysis of co-occurrence tensor. 
\end{enumerate}

\subsection{Construction of initial corpus}
The main dataset of this study is news articles from four major U.S. news agencies --- \textit{New York Times}, \textit{Washington Post}, \textit{Wall Street Journal}, and \textit{Fox News} --- published between January 1, 2018 and June 16, 2018. We first scraped all news articles containing ``Korea'' between January 1, 2018 and June 16, 2018 from the websites of four major U.S. news agencies. Then, we performed elementary pre-processing operations on the scraped articles. For instance, we excluded irrelevant contents such as advertisements, photographs, book reviews, and video clips without textual information. We also removed duplicate news articles posted more than once, only keeping the original articles. There remained 2,951 news articles after pre-processing.  


\subsection{Construction of co-occurrence tensor}
Using the initial corpus of Korea-related news articles, we construct a co-occurrence tensor of key figures. Unlike citation network analysis \citep{Price1965, Newman2001} in which node sets are easily identifiable, identification of ``key figures" is one of the most challenging tasks, considering that researchers in general do not know who the key figures are in a particular issue. We tried to solve this problem by first identifying a name in an article, factoring in the variant spellings of names, the absolute frequency and the frequency over the time-period of study. To identify a name, we used a list of U.S. surnames provided by the U.S. Census Bureau (\url{https://www2.census.gov}) and a list of Korean surnames from Statistics Korea (\url{http://kostat.go.kr}) to identify people's names in the initial corpus. We paid close attention to spelling variations arising from anglicization, typos, and the presence or the absence of middle names and made appropriate adjustments.\footnote{For example, North Korean leader Kim Jong-un appears as ``Kim Jong Un'',``Kim Jung Un'', ``Kim Jung Eun'',  ``Mr.\ Kim,'' or ``Leader Kim.''} We considered a name a part of ``key figures" if it was mentioned at least ten times in the corpus. We further put a threshold on names based on the frequency consistency. That is, we chose the names that show up at least a quarter of the entire period. In the end, there remained 34 names of key figures of the U.S.-North Korea relations.

\begin{figure}
        \centering
        \includegraphics[scale = 0.45]{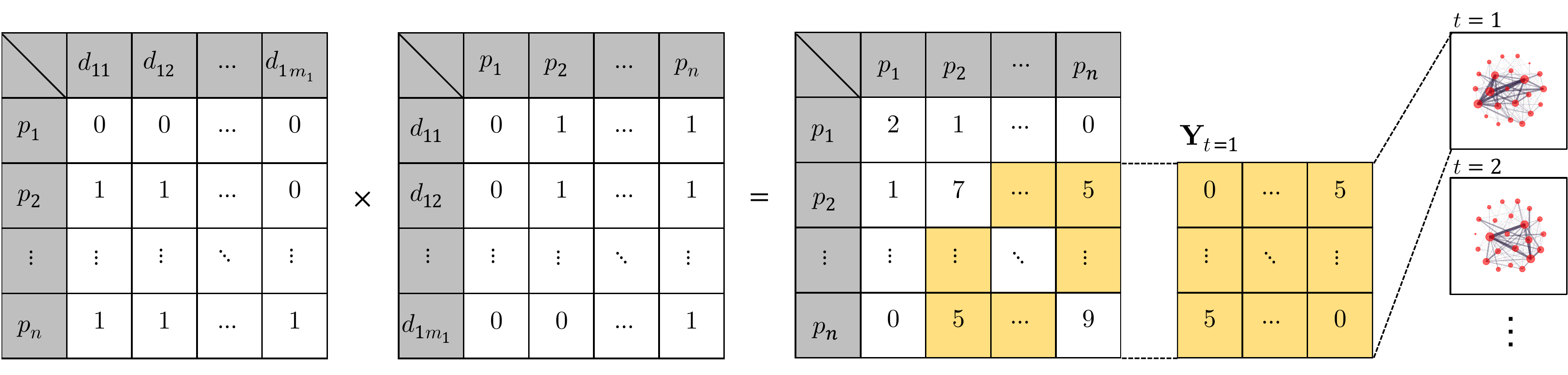}
        \caption{\emph{Structure of key figure network data}}
        \label{fig:data_network}
    \end{figure}

The next step was to construct an $N \times N \times T$ co-occurrence tensor of the 34 key figures. Given the six-month window, we decided to aggregate information by week. We counted the frequency of co-occurrence of the names of the key figures on weekly basis. Figure \ref{fig:data_network} shows the structure of the key figure network data, where $p_{i}$, $d_{tj}$, $m_t$, $\Y_t$ denote, repectively, the $i$th potential key figure, the $j$th article of the $t$th week,the number of articles in $t$\textsuperscript{th} week, and a people co-occurrence matrix at week $t$.

The first matrix in Figure \ref{fig:data_network} indicates key figure-article matrix having each key figure in each row and each article in each column. Note that each entry in the matrix takes either $0$ (\textit{no-occurrence}) or $1$ (\textit{occurrence}).  The second matrix is a transpose of the first matrix. Multiplying these two matrices produces a co-occurrence matrix of all the key figures. From this matrix, we extract the co-occurrence of the selected key figures and construct $\Y_t$.

 %


\subsection{Bayesian hidden Markov multilinear tensor analysis}
 \cite{Sohn2017} and \cite{Park2017} presented a Bayesian hidden Markov multilinear tensor method (HMTM) for analyzing block structure changes in longitudinal network data. The method extends \cite{Hoff2015}'s Bayesian multilinear tensor model  into change-point analysis. The HMTM first corrects degree information in longitudinal network data in order to filter confounding information of the baseline expectation of associations among pairs of nodes when the quantity of our interest lies in the discovery of latent block structures  \citep{peixoto2013eigenvalue}. For degree-correction, we use a principal eigen matrix $\mathbf{\Omega}_t$, consisting of the principal eigenvalue ($\lambda^{princ}_t=\text{max}(\vert\lambda(\mathbf{Y}_t)\vert)$) and its associated eigenvector:  $\mathbf{B}_{t} = \mathbf{Y}_t - \mathbf{\Omega}_t$. Now, let $S_t$  denote a hidden state variable ($S_t \in \{1, \ldots, k+1\}$ for $t = 1, \ldots, T$) and $\P$ denote  a $k+1$ by $k+1$ transition matrix where $p_{i,i}$ is the $i$th diagonal element of $\P$. Then, a HMTM with $k$ breaks is 
\begin{eqnarray}\label{eq:cp}\label{eq1}
	\mathbf{B}_{t} &=& \U_{S_t}\mathbf{V}_t\U_{S_t}^T + \mathbf{E}_t\\\label{eq3}
	S_t|S_{t-1}, \P &\sim&\mathcal{M}arkov(\P, \pi_0)\\\label{eq4}
	p_{i,i}&\sim& \betadist{a_0}{b_0}\\\label{eq5}
	\mathbf{E}_t &\sim&  
\mathcal{N}_{N \times N}(\mathbf{0}, \sigma_{S_t}^2\mathbf{I}_{N}, \mathbf{I}_N) 
\end{eqnarray}
where $\pi_0$ is the initial probability ($\pi_0 = (1, 0, \ldots, 0)$) and $\betadist{.}{.}$ is a Beta distribution. The core part of the HMTM is the decomposition of longitudinal network data into regime-dependent $\U_{S_t}\mathbf{V}_t\U_{S_t}^T$ where $\U_{S_t}$ represents regime-dependent latent node positions and $\mathbf{V}_t$ represents time-varying network generation rules.\footnote{The algorithms are available at \texttt{NetworkChange} \citep{NetworkChange}, which is  an open-source \texttt{R} package. A brief discussion of HMTM is available in Supplementary Material. }

\section{Results}

\begin{table}[ht]\footnotesize
\centering
\begin{tabularx}{\textwidth}{rlcX}\toprule
\toprule
&Name &Date  & Summary \\
\midrule
Regime 1 & \emph{Nuclear Button} & 01-01$\sim$02-04& Highest tension between U.S. and North Korea over a possible preemptive strike against North Korea's nuclear facilities \\
Regime 2 & \emph{Winter Olympic} & 02-05$\sim$03-04& The Winter Olympics in South Korea and North Korea's participation  \\
Regime 3 & \emph{Summit Announcement} & 03-05$\sim$03-11&  President Trump's acceptance of Kim Jong-un's summit letter \\ 
Regime 4 & \emph{Appointment of Pompeo} & 03-12$\sim$03-18& President Trump appoints Mike Pompeo as the Secretary of State \\
Regime 5 & \emph{High-level Talks} &03-19$\sim$06-03& U.S.-North Korea High-level Talks\\ 
Regime 6 & \emph{Summit Eve} &06-04$\sim$06-10& Working-level discussion at Panmunjom for U.S.-North Korea summit\\ 
Regime 7 & \emph{Singapore Summit} &06-11$\sim$06-17& U.S.-North Korea summit in Singapore \\
\bottomrule
\end{tabularx}
   \caption{\emph{Descriptive Summary of the Seven Regimes}}\label{regime}
\end{table}

Table \ref{regime} reports a descriptive summary of seven regimes, which is in a chronological order of events, uncovered by our change-point analysis.\footnote{It turns out that the six break model is the most reasonable choice given the data and the chosen model. Readers are advised to consult the Supplementary Material for detailed discussions on break number detection using log marginal likelihoods regarding this choice.} The change-point analysis captures important turning points of the U.S.-North Korea relations during the first two quarters of 2018. Based on our reading of the case, we label each regime \textit{Nuclear Button}, \textit{Winter Olympic}, \textit{Summit Announcement}, \textit{Appointment of Pompeo}, \textit{High-level Talks}, \textit{Summit Eve}, and \textit{Singapore Summit}.  

Regime 1 corresponds to the period of the highest tension between two countries. During this period, the Trump administration was allegedly considering a preventive attack on North Korea's nuclear facilities (so-called bloody nose operation). Regime 2 points to a period of the Winter Olympics during which North Korea attempted several rounds of dialogues with South Korea and the U.S. Regime 3 covers a period of the warm relationship between North Korea and the U.S., which started from Kim Jong-un's letter which suggested a summit with President Trump. Regime 4 represents the period when President Trump fired Rex Tillerson and replaced him as Mike Pompeo. Regime 5 covers high-level talks between U.S.and North Korea. Regime 6 corresponds to working-level discussion between U.S. and North Korea regarding summit. Finally, Regime 7 covers the week of the U.S.-North Korea summit in Singapore.

   \begin{figure}
        \centering
        \includegraphics[scale = 0.5]{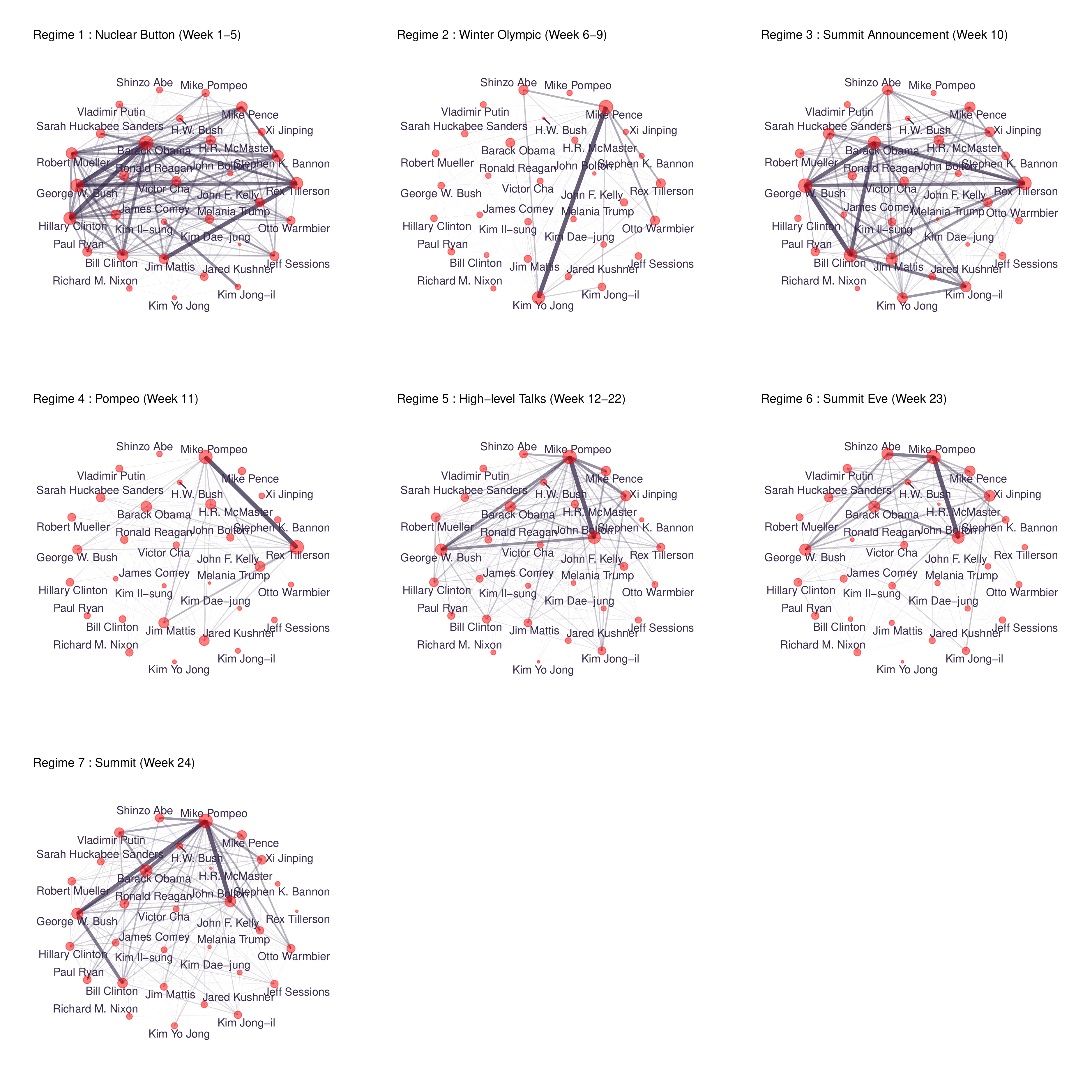}
        \caption{\emph{Regime Changes in Key Figure Networks in the 2018 US-North Korea Relationship}}
        \label{fig:regime}
    \end{figure}


Figure \ref{fig:regime} shows the network topology of regime changes in the key figure networks. We highlight several interesting patterns. 

First, \texttt{Jim Mattis}, \texttt{Rex Tillerson}, and \texttt{H.R. McMaster} very often co-occur during the \textit{Nuclear Button} regime. During this period, \texttt{H.R. McMaster} was reported to be one of the most active proponents of a preventive attack on North Korea -- so-called ``bloody nose operation'' -- while \texttt{Rex Tillerson} and  \texttt{Jim Mattis} allegedly opposed a preventive attack on North Korea within the National Security Council. 

The second notable pattern in Figure \ref{fig:regime} is the co-occurrence network of U.S.\ Vice president \texttt{Mike Pence} and \texttt{Kim Yo Jong} during Regime 2, with the latter being the younger sister of Kim Jong-un who visited the 2018 Winter Olympics opening ceremony in Pyeongchang, South Korea as a special envoy of Kim Jong-un.  \texttt{Mike Pence} also visited the 2018 Winter Olympics and there was a speculation that they might hold a meeting during their stay in South Korea, which reportedly did not happen. 

The third notable pattern is that names of former U.S. presidents (ex. \texttt{George W.\ Bush}, \texttt{Barack Obama}, and \texttt{Bill Clinton}) very often co-occur. During the sample period, many news articles compared the Trump administration's approach to North Korea with the approaches of previous U.S. administrations.

A strong link between \texttt{Mike Pompeo} and \texttt{Rex Tillerson} during Regime 4 was due to the taking office of \texttt{Mike Pompeo} for the Secretary of State replacing \texttt{Rex Tillerson}. From then on, \texttt{Mike Pompeo} has become a central figure in the U.S.-North Korea relationship. \texttt{John Bolton} who replaced \texttt{H.R. McMaster} as national security adviser shows a similar pattern with \texttt{H.R. McMaster} but the number of their co-appearances is much smaller than the duo of Pompeo-Tillerson. 

The center and right panels in the second row of Figure \ref{fig:regime} (Regime 5 and Regime 6) show that \texttt{Xi Jinping} became more important during these periods. Since Kim Jong-un's surprise visit to China on March 11, President Trump had complained that China interfered in U.S.-North Korea talks.\footnote{For example, on May 18, 2018, President Trump said ``It could very well be that he[Xi Jinping]'s influencing Kim Jong-un. We'll see what happens." \emph{South China Morning Post}, ``Donald Trump says North Korea could have been `influenced' by Xi Jinping to turn on US ahead of talks with Kim Jong-un'' May 18, 2018.} Various sources reported that Trump canceled the summit to send North Korea a message that China should not meddle with the US-North Korea dialogue.

Lastly, \texttt{Robert Mueller}'s strong links to other key figures in Regime 1 are due to the co-occurrence of the U.S.-North Korea issues with the Russia investigation by Special Counsel Robert Mueller.  However, the two issues  appear separately in news articles after Regime 1 as both issues became important issues. Accordingly, \texttt{Robert Mueller}'s links to other key figures became weaker after Regime 1.

\section{Conclusion}
In this paper, we presented a simple method for analyzing a dynamic network of key figures in the U.S.-North Korea relations by taking co-occurrence of people's names in an article as a domain-relevant social link. We called a group of people that co-occur repeatedly in the same domain the ``key figures'' and their corresponding social networks the ``key figure networks.''  Then, we analyzed block-structure changes of the key figure networks in the U.S.-North Korea relations using a Bayesian hidden Markov multilinear tensor model. 

The results of our analysis showed that block structure changes in the key figure network in the U.S.-North Korea relations are highly consistent with important turning points in the U.S.-North Korea relations during the first two quarters of 2018. We also found that the rise and fall of key figures in the key figure network closely reflect the fluctuations of their importance or relevance in each phase. Our method can be applied to other text data that contains information of latent social links among important political actors. One example that we have been investigating is the UN General Debate Corpus. This data contains speeches of country representatives that address important international or country-specific issues. 

There are many directions for future developments of our method. One direction is to distinguish the \emph{nature} of co-occurrence links. For example, the nature of co-occurrence links could be positive, negative, or directional. Also, we could build a bipartite network between terms (or phrases) and key figures. Some key figures appear more frequently with certain terms than others. Advanced text analysis methods such as word embeddings \citep{pennington2014glove} or structural topic model \citep{Roberts2016} combined with various network analysis methods will allow us to uncover important latent social networks from political text data. 
\end{spacing}

\clearpage
\appendix 
\section*{News Articles and Key Figures}
Table \ref{fig:key} shows an example of news articles and key figures mentioned in the article. 
\begin{table}[h]\footnotesize
\begin{tabularx}{\textwidth}{l l X X}\toprule
Source &Date & Content &Figures mentioned in the article \\\midrule
Fox News & 2018.1.2 & \tiny North Korea team at Olympics should prompt US boycott, \texttt{Graham} says. South Korea has offered to hold high-level talks with rival North Korea next week \ldots. Tuesday's offer came a day after North Korean leader  \texttt{Kim Jong Un} said in his New Year's address that he was willing to send a delegation to the Pyeongchang Winter Olympics, \ldots. But the prospect of a North Korean team at the Winter Games drew a blistering response Monday from U.S. Sen.\ \texttt{Lindsey Graham}, R-S.C., who proposed that if North Korea attends the Olympics, then the United States should not. \ldots In 2014,  \texttt{Graham} also pitched a boycott of the Winter Olympics in Sochi, Russia, objecting then to the prospect of Moscow offering asylum to NSA whistleblower  \texttt{Edward Snowden} \ldots \texttt{Sung-Yoon Lee}, a Tufts University Fletcher School professor and North Korea expert, told the Boston Herald that over the last 25 years the U.S. and its allies ``have a less well-developed game plan and no real strategy.'' &\texttt{Lindsey Graham}, \texttt{Kim Jong Un},  \texttt{Edward Snowden}, \texttt{Sung-Yoon Lee}\\
New York Times & 2018.1.2 & \tiny
North Korea's leader, \texttt{Kim Jong-un}, had suggested on Monday that the countries open dialogue on easing military tensions \ldots President \texttt{Trump} responded somewhat cautiously to \texttt{Mr. Kim}'s overture early Tuesday, \ldots  \texttt{Mr. Trump}, referring to \texttt{Mr. Kim}, said: ``Will someone from his depleted and food starved regime please inform him that I too have a Nuclear Button, but it is a much bigger \& more powerful one than his, and my Button works!" \ldots Speaking at the United Nations on Tuesday, the United States Ambassador, \texttt{Nikki R. Haley}, appeared to dismiss the potential for bilateral negotiations between North and South Korea. \ldots \texttt{Heather Nauert}, the State Department spokeswoman, said the Trump administration was still assessing whether the United States supported direct talks between South Korea and North Korea that excluded the United States.
\texttt{Cho Myoung-gyon}, the South's point man on the North, \ldots  since President \texttt{Moon Jae-in}'s conservative predecessor, the impeached President \texttt{Park Geun-hye}, shut down a joint industrial complex in the North Korean town of Kaesong in early 2016.
&\texttt{Kim Jong Un}, \texttt{Trump}, \texttt{Nikki R. Haley}, \texttt{Heather Nauert}, \texttt{Moon Jae-in}, \texttt{Park Geun-hye}, \texttt{Cho Myoung-gyon}\\
\bottomrule
\end{tabularx}
\caption{\emph{Examples of News Articles Containing Key Figures of the US-North Korea Dialogue}}\label{fig:key}
\end{table}

\begin{figure}
        \centering
        \includegraphics[scale = 0.69]{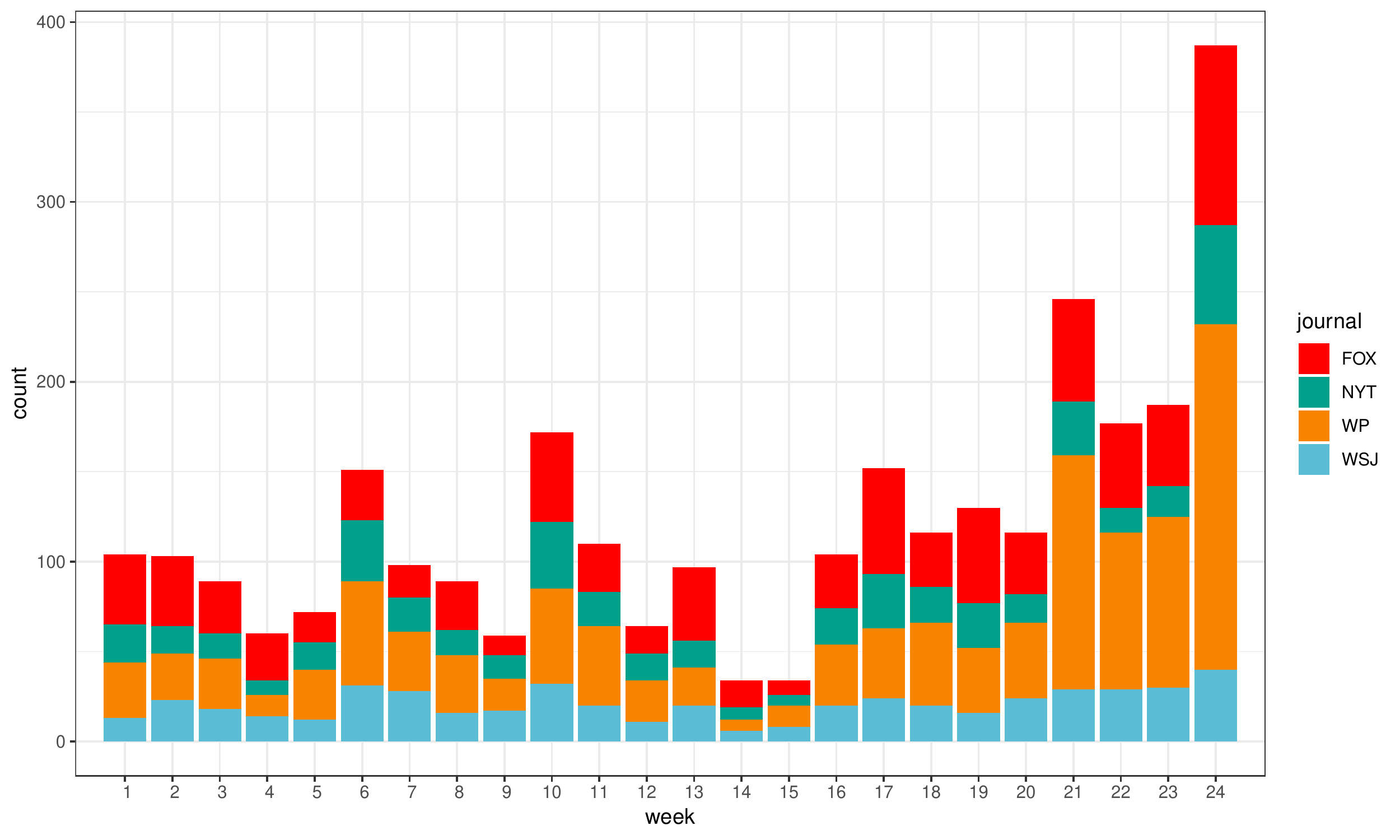}
        \caption{\emph{Summary of filtered news articles between January 1, 2018 and June 16, 2018}}
        \label{fig:summary}
    \end{figure}

A  summary of the pre-processed data is shown in Figure \ref{fig:summary}. 

\clearpage
\section*{Descriptive Information of Key Figure Networks}
\begin{figure}[h]
        \centering
        \includegraphics[scale = 0.43]{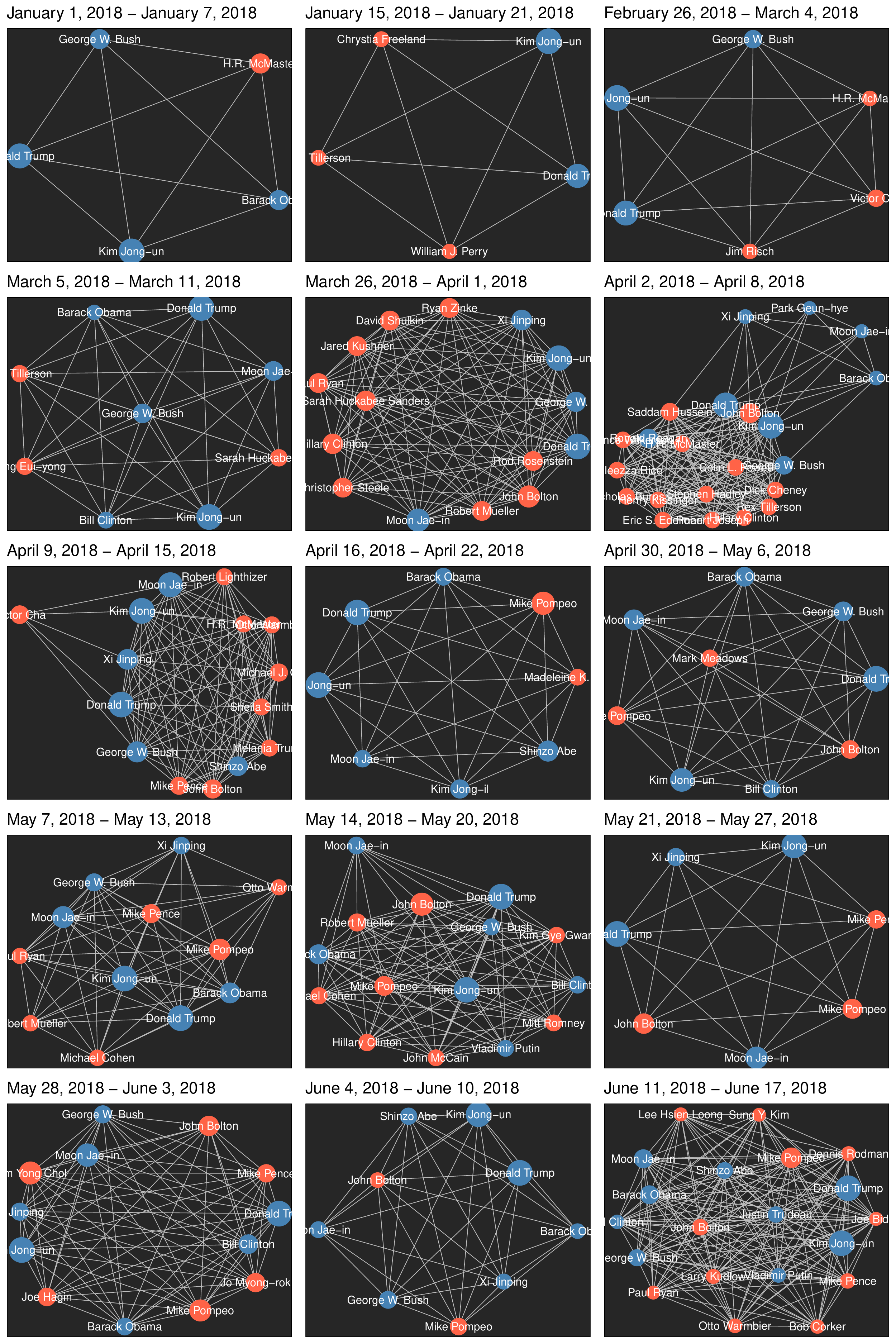}
        \caption{\emph{Key figure network over time}}
        \label{fig:keyfigure}
    \end{figure}

Figure \ref{fig:keyfigure} displays key figure networks from January 1, 2018 to June 17, 2018 by aggregating key figure networks on a weekly basis. We chose 15 weeks for better display in the paper. 

\begin{figure}
        \centering
        \includegraphics[scale = 0.56]{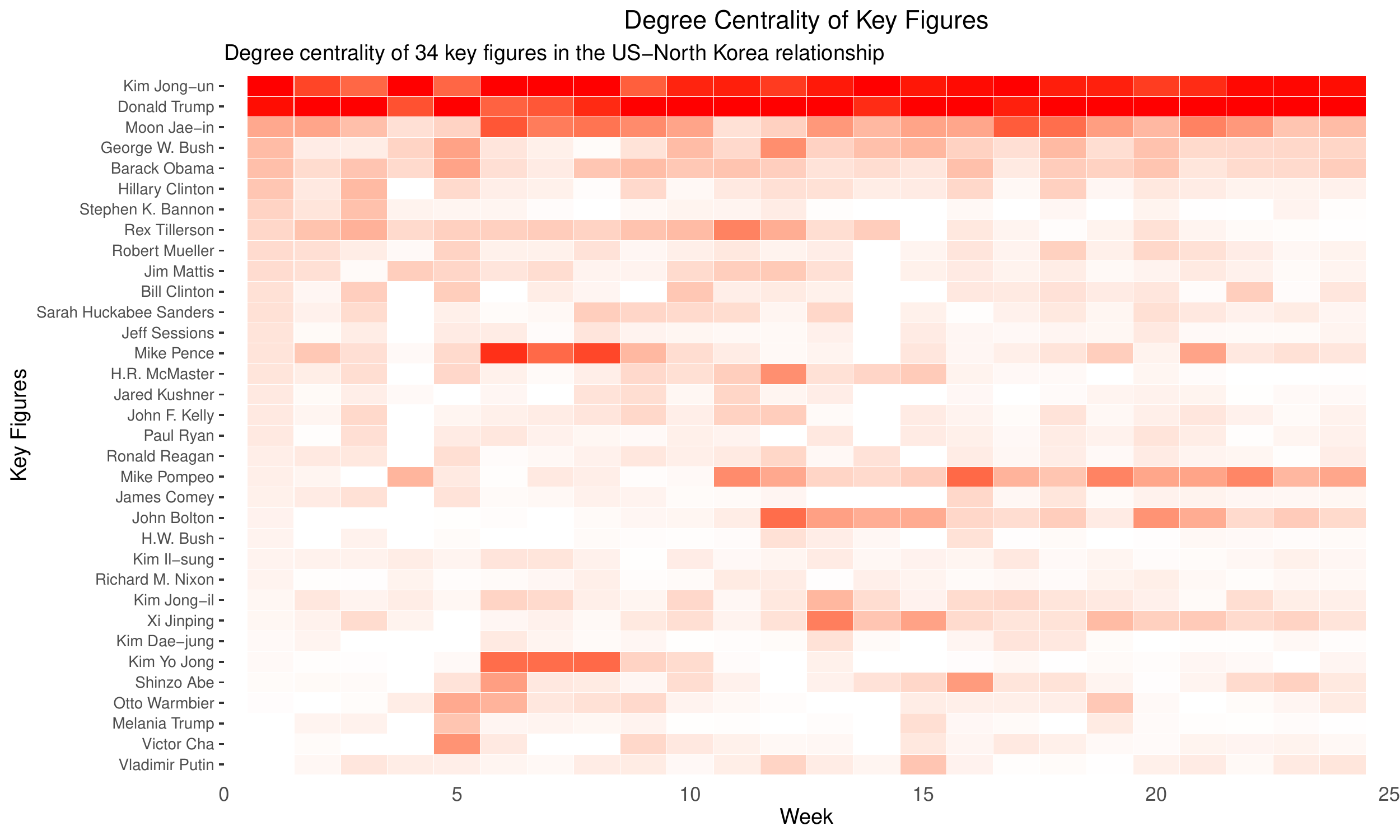}
        \caption{\emph{Degree Centrality of Key Figure Networks}}
        \label{fig:deg}
    \end{figure}

\begin{figure}[h]
        \centering
        \includegraphics[scale = 0.65]{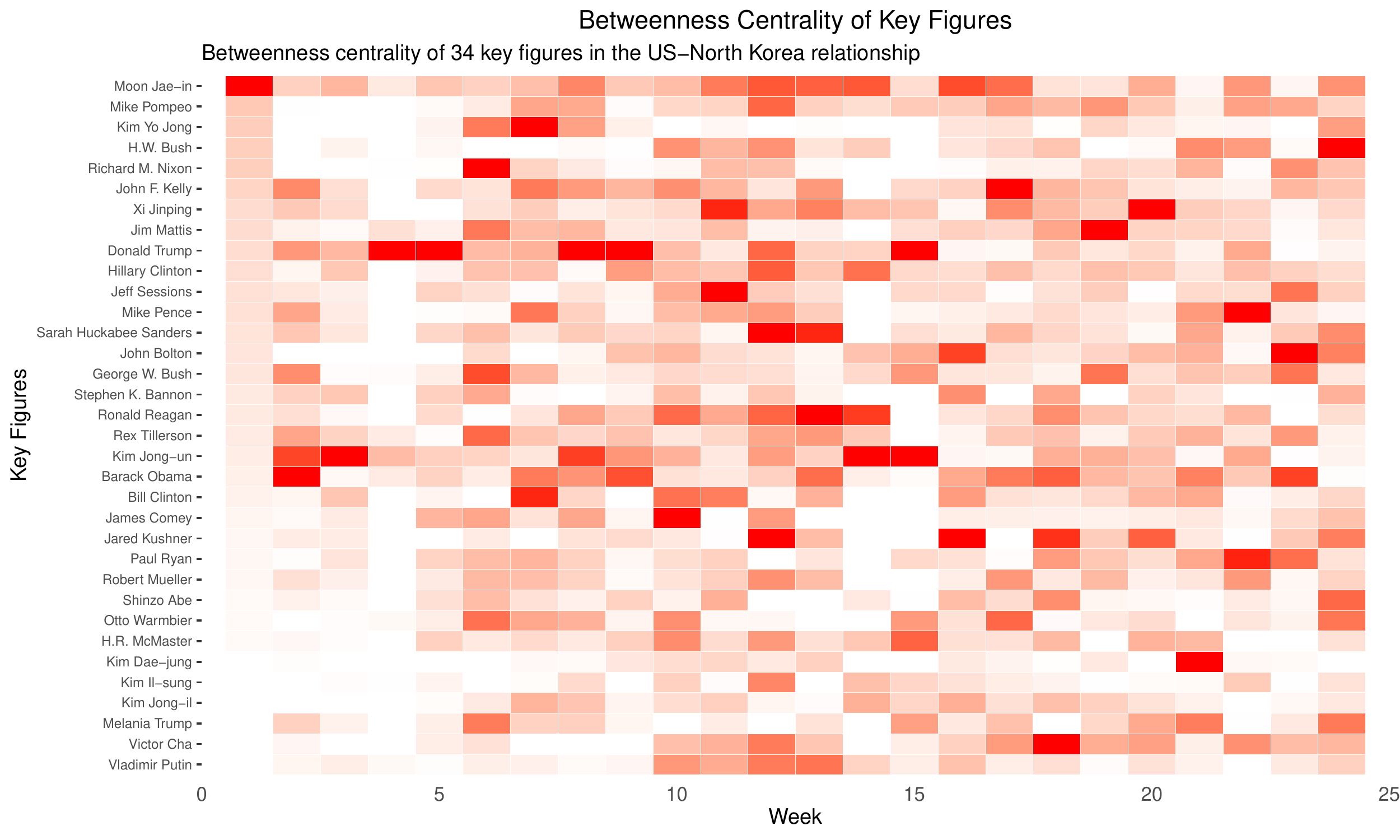}
        \caption{\emph{Betweenness Centrality of Key Figure Networks}}
        \label{fig:bw}
    \end{figure}

Figure \ref{fig:deg} shows the degree centrality scores of the 34 key figures in the U.S.-North Korea relationship. Dark colors indicate high degree centrality scores with light colors indicating the opposite. This clearly shows \emph{the rise and fall} of some key figures. For example, the degree centrality scores of Rex Tillerson and H.R. McMaster dropped after Week 14 and Week 15 while the degree centrality scores of John Bolton and Mike Pompeo increased after Week 10 and Week 11. However, the degree centrality does not show relational information (i.e. who is connected with whom). In other words, changes in the key figure networks of the U.S.-North Korea relationship must have occurred between other key figures. As such, we drop three key figures (namely, Kim Jong-un, Donald Trump, and Moon Jae-in) from the data for the change-point analysis since Kim Jong-un, Donald Trump, and Moon Jae-in have high degree centrality scores throughout the entire sample period. Figure \ref{fig:bw} shows betweenness centrality of key figure networks across 24 weeks.


Table \ref{degcen} summarizes top 100 of key figures based on the total number of news articles that mentioned each figures.  


\begin{table}[ht]\tiny
\centering
\begin{tabular}{lllllllr}
\toprule
Name & Degree & Name & Degree& Name & Degree& Name & Degree\\
\midrule
Kim Jong-un & 2447 & Ronald Reagan & 96 & Gina Haspel & 48 & Cory Gardner &  34 \\ 
  Donald Trump & 2394 & Ivanka Trump & 90 & Rod Rosenstein & 48 & Joe Hagin &  34 \\ 
  Moon Jae-in & 1025 & Chung Eui-yong & 85 & Nancy Pelosi & 47 & Michael J. Green &  33 \\ 
  Mike Pompeo & 636 & Paul Ryan & 85 & Stephen K. Bannon & 47 & Robert Menendez &  33 \\ 
  Mike Pence & 387 & Justin Trudeau & 82 & Christopher R. Hill & 46 & Wilbur Ross &  33 \\ 
  John Bolton & 353 & Richard M. Nixon & 73 & Larry Kudlow & 46 & Peter Navarro &  32 \\ 
  Barack Obama & 339 & Rudolph W. Giuliani & 72 & Marco Rubio & 46 & Suh Hoon &  32 \\ 
  Xi Jinping & 339 & Kim Dae-jung & 71 & Roh Moo-hyun & 46 & Kim Dong Chul &  31 \\ 
  George W. Bush & 324 & Michael Cohen & 71 & Choe Son-hui & 44 & Kim Sang-deok &  31 \\ 
  Shinzo Abe & 241 & James Comey & 64 & John McCain & 44 & Ri Son Gwon &  31 \\ 
  Rex Tillerson & 240 & Chuck Schumer & 63 & John Roberts & 41 & Jeffrey Lewis &  30 \\ 
  Kim Jong-il & 228 & Dennis Rodman & 63 & Kang Kyung-wha & 41 & John Delury &  30 \\ 
  Kim Yo Jong & 195 & Steven Mnuchin & 63 & Sung Y. Kim & 40 & Michael Flynn &  30 \\ 
  Otto Warmbier & 162 & Mitch McConnell & 62 & Melania Trump & 39 & Adam Schiff &  29 \\ 
  Kim Yong Chol & 160 & Joseph Yun & 60 & Park Geun-hye & 39 & Kim Jong Nam &  29 \\ 
  Jim Mattis & 157 & Lindsey O. Graham & 59 & H.W. Bush & 38 & Rand Paul &  29 \\ 
  Bill Clinton & 144 & Kim Yong Nam & 57 & Madeleine K. Albright & 38 & Taro Kono &  29 \\ 
  Sarah Huckabee Sanders & 142 & Emmanuel Macron & 56 & Cho Myoung-gyon & 37 & Yoon Young-chan &  29 \\ 
  Kim Il-sung & 131 & Heather Nauert & 55 & Saddam Hussein & 37 & Ma Ying-jeou &  28 \\ 
  Robert Mueller & 128 & Jeff Sessions & 54 & Ji Seong-ho & 36 & Joel Wit &  27 \\ 
  Hillary Clinton & 127 & Jared Kushner & 52 & Ri Sol Ju & 36 & Kim Hak Song &  27 \\ 
  Vladimir Putin & 126 & Angela Merkel & 50 & Ri Yong Ho & 36 & Bruce Klingner &  26 \\ 
  H.R. McMaster & 107 & Bob Corker & 50 & Bernie Sanders & 35 & Foster Klug &  26 \\ 
  John F. Kelly & 106 & Jimmy Carter & 50 & Kim Eui-kyeom & 35 & Harry Harris &  26 \\ 
  Victor Cha & 106 & Nikki R. Haley & 49 & Scott Pruitt & 35 & Jo Myong-rok &  26 \\ 
   &  &  &  &  &  & Robert Lighthizer &  26 \\ 
   &  &  &  &  &  & Sue Mi Terry &  26 \\ 
\bottomrule
\end{tabular}
  \caption{\emph{Top 100 of Key Figures mentioned in news articles across 24 weeks}}\label{degcen}
\end{table}
 
\clearpage
\section*{A Brief Introduction of Hidden Markov Tensor Model}
Tensor decomposition is becoming a standard means to analyze longitudinal network datasets \citep{hoff2009_cmot, Hoff2011, Rai2015, Hoff2015, Minhas2016, johndrow2017, Han2018MultiresolutionTD}. A longitudinal network data set can be represented as a tensor $\mathcal{Y}=\{\Y_t\vert t\in\{1,\ldots,T\}\}\in\real^{N \times N \times T}$, which is an array of $N \times N$ square matrices $\Y_t=\{y_{ijt}\vert i,j\in\{1,\ldots,N\}\}$. Here $y_{ijt}$ informs the dyadic relationship between actors $i$ and $j$ at time $t$. 

HMTM is a dynamic network model using \cite{Hoff2011, Hoff2015}'s multilinear tensor regression model (MTRM), which is a multilayer (\emph{i.e.} tensor) extension of the latent space approach to network data. MTRM allows us to decompose longitudinal network data into node-specific (or row and column) random effects and time-specific (or layer-specific) random effects.

Following the MTRM, the HMTM decomposes the degree corrected network data at $t$ as a bilinear product of latent node positions and dimension weights subject to hidden state changes: 
\begin{eqnarray}\label{hmtm1}
	\mathbf{B}_{t} &=&\U_{S_t} \mathbf{V}_t  \U_{S_t}^T + \mathbf{E}_t\\\label{hmtm2}
	\mathbf{E}_t &\sim&  \left\{ \begin{array}{rcl}\label{hmtm3}
\mathcal{N}_{N \times N}(\mathbf{0}, \sigma_{S_t}^2\mathbf{I}_{N}, \mathbf{I}_N) &\mbox{for}  & \text{Normal Error} \\ 
\mathcal{N}_{N \times N}(\mathbf{0}, \gamma_t^{-1}\sigma_{S_t}^2\mathbf{I}_{N}, \mathbf{I}_N) & \mbox{for} & \text{Student-$t$ Error}
\end{array}\right.
\end{eqnarray}

For prior distributions of $\U$ and $\V$, we follow \cite{Hoff2011}'s hierarchical scheme with two major  modifications.  First, we orthogonalize each column of $\mathbf{U}_{S_t}$ using the Gram-Schmidt process \citep{Bjorck1996, Dunson2015} in each simulation step. \cite{Hoff2011}'s hierarchical scheme centers rows of $\mathbf{U}_{S_t}$ around its global mean ($\boldsymbol{\mu}_{u,S_t}$) using a multivariate normal distribution. This does not guarantee the orthogonality of each latent factor in $\mathbf{U}_{S_t}$. The lack of orthogonality makes the model unidentified, causing numerical instability in parameter estimation and model diagnostics \citep{murphy2012machine, Dunson2015}.%

Second, we use independent inverse-gamma distributions instead of inverse-Wishart distribution for the prior distribution of a variance parameter ($\Psi_{u, S_t}, \Psi_{v}$). The use of inverse-Wishart distribution for the prior distribution of a variance parameter ($\Psi_{u, S_t}, \Psi_{v}$) comes at a great cost because choosing informative inverse-Wishart prior distributions for $\Psi_{u, m}$ and $\Psi_{v}$ is not easy \citep{Gelman2015} and a poorly specified inverse-Wishart prior distribution has serious impacts on the marginal likelihood estimation. In our trials, the log posterior  inverse-Wishart density  of $\Psi_{u, S_t}$ and $\Psi_{v}$ often goes to a negative infinity, failing to impose proper penalties. In HMTM, the off-diagonal covariance of $\mathbf{U}_m$ is constrained to be 0, thanks to the Gram-Schmidt process, and the off-diagonal covariance of $\mathbf{V}$ is close to 0 as $\v_t$ measures time-varying weights of independent $\mathbf{U}_m$. Thus, inverse-gamma distributions resolve a computational issue without a loss of information. 
 
The resulting prior distributions of $\U$ and $\V$ are matrix-variate normal distributions in which each column vector ($ \u_{i, S_t}$ and $\v_{t}$) follows a multivariate normal distribution. We first discuss the prior distribution of $\U$: 
\begin{eqnarray}\label{prior1}
\U_{S_t} &\equiv&  (\u_{1, S_t}, \ldots, \u_{N, S_t})^\top\in\mathbb{R}^{N\times R}\\\label{prior2}
 \u_{i, S_t}&\sim& \mathcal{N}_{R}(\boldsymbol{\mu}_{u,S_t}, \Psi_{u, S_t})\\\label{prior3}
\boldsymbol{\mu}_{u, S_t}|\Psi_{u, S_t} &\sim& \mathcal{N}_{R}(\boldsymbol{\mu}_{0, u_{S_t}}, \Psi_{u, S_t})\\ \label{prior4}
\Psi_{u, S_t} &\equiv& \left( \begin{array}{ccc}
 \psi_{1, u, S_t} & \ldots & 0 \\
0 & \psi_{r, u, S_t} & 0 \\
0 &  \ldots & \psi_{R, u, S_t} \end{array} \right)\\\label{prior5}
	\psi_{r, u, S_t} &\sim&\mathcal{IG}\left (\frac{u_0}{2}, \frac{u_{1}}{2}\right).
\end{eqnarray}
	
The prior distributions of $\V$ are similar to $\U$ but one difference is that only diagonal elements of $\V_t$ are modeled as a multivariate normal distribution:   	
\begin{eqnarray}\label{prior6}
\mathbf{V}_t &\equiv&  \left( \begin{array}{ccc}
 v_{1, t} & \ldots & 0 \\
0 & v_{r, t} & 0 \\
0 &  \ldots & v_{R, t} \end{array} \right) \\\label{prior7}
\v_{t} &\equiv&(v_{1, t}, \ldots, v_{R, t})^\top\in\mathbb{R}^{R\times 1} \\
\v_{t}&\sim& \mathcal{N}_{R}(\boldsymbol{\mu}_v, \Psi_v)\\\label{prior8}
\boldsymbol{\mu}_{v}|\Psi_{v} &\sim& \mathcal{N}_{R}(\boldsymbol{\mu}_{0, v}, \Psi_{v})\\ \label{prior9}
\Psi_{v} &=& \left( \begin{array}{ccc}
\psi_{1, v} & \ldots & 0 \\
0 & \psi_{r, v} & 0 \\
0 &  \ldots & \psi_{R, v} \end{array} \right)\\\label{prior10}
	\psi_{r, v} &\sim&\mathcal{IG}\left (\frac{v_0}{2}, \frac{v_{1}}{2}\right).
\end{eqnarray}

Then, we complete the model building by introducing HMM-related prior specifications following \cite{Chib1998}: 
\begin{eqnarray}\label{eq:cp1}
	S_t|S_{t-1}, \P &\sim&\mathcal{M}arkov(\P, \pi_0)\\\label{eq:cp10}
	\underbrace{\P}_{M \times M} &=& (\mathbf{p}_1, \ldots, \mathbf{p}_M)\\\label{eq:cp11}
	\mathbf{p}_i &\sim& \text{Dirichlet}(\alpha_{i,1}, \ldots, \alpha_{i,M})
\end{eqnarray}
where $\pi_0$ is the initial probability of a non-ergodic Markov chain ($\pi_0 = (1, 0, \ldots, 0)$).

\subsection*{MCMC Algorithm for Hidden Markov Tensor Model}
\begin{description}
\item[] For each $t$ layer, generate $\mathbf{B}_t = \mathbf{Y}_t - \mathbf{\Omega}_t$ by choosing a null model ($\mathbf{\Omega}_t$).
\item[] Set the total number of changepoints $M$ and initialize ($\mathbf{U}, \boldsymbol{\mu}_u, \Psi_v, \V, \boldsymbol{\mu}_v, \Psi_v, \beta, \sigma^2, \S, \P)$.
\end{description}

\noindent \textbf{Part 1}

\begin{description}\footnotesize
\item[Step 1] The sampling of regime specific $\U, \boldsymbol{\mu}, \Psi_{u}$ consists of the following three steps for each regime $m$.  Let $ \Psi_{u} = \left( \begin{array}{ccc}
\psi_{1, u, m} & \ldots & 0 \\
0 & \psi_{r, u, m} & 0 \\
0 &  \ldots & \psi_{R, u, m} \end{array} \right) $.
	\begin{enumerate}
	\item $p(\psi_{r, u, m} |  \B,  \P, \S, \mathbf{\Theta} ^{-\Psi_{u, m}}) \propto \mathcal{IG}\left (\frac{u_0 + N}{2}, \frac{\U_{r, m}^T\U_{r, m} + u_{1}}{2}\right)$.
	
	\item $p(\boldsymbol{\mu}_{u, m}|  \B,  \P, \S, \mathbf{\Theta} ^{-\boldsymbol{\mu}_{u, m}}) \propto  \textrm{multivariate normal}(\U_m^T\mathbf{1}/(N + 1), \Psi_{u, m}/(N + 1))$.
	\item $p(\U_m|  \B,  \P, \S, \mathbf{\Theta} ^{-\U_m}) \propto \text{matrix normal}_{N \times R}(\tilde{\M}_{u, m}, \mathbf{I}_{N}, \tilde{\Psi}_{u, m})$ 
	where 
	\begin{eqnarray*}
	\tilde{\Psi}_{u, m} &=& (\mathbf{Q}_{u, m}/\sigma_m^2 + \Psi_{u, m}^{-1})^{-1} \\
	\tilde{\M}_{u, m} &=& (\mathbf{L}_{u, m}/\sigma_m^2 + \mathbf{1}\boldsymbol{\mu}_{u, m}^T \Psi_{u, m}^{-1}) \tilde{\Psi}_{u, m}\\
	\mathbf{Q}_{u, m} &=& (\U_m^T\U_m)\circ(\V_m^T\V_m) \\
	\mathbf{L}_{u, m} &=& \sum_{j, t:\; t \in S_t = m}b_{\cdot,j, t} \otimes (\U_{m, j, \cdot} \circ   \V_{m, t, \cdot} )
	\end{eqnarray*}
	\item Orthogonalize $\U_m$ using the Gram-Schmidt  algorithm. 
	\end{enumerate}

\item[Step 2] The sampling of $\V, \boldsymbol{\mu}_v, \Psi_v$ is done for each regime.  Let $ \Psi_{v} = \left( \begin{array}{ccc}
\psi_{1, v, m} & \ldots & 0 \\
0 & \psi_{r, v, m} & 0 \\
0 &  \ldots & \psi_{R, v, m} \end{array} \right) $.
	\begin{enumerate}
	\item $p(\psi_{r, v, m} |  \B,  \P, \S, \mathbf{\Theta} ^{-\Psi_{v, m}}) \propto\mathcal{IG}\left (\frac{v_0 + T}{2}, \frac{\V_{r, m}^T\V_{r, m} + v_{1}}{2} \right)$.
	\item  $p(\boldsymbol{\mu}_{v, m}|  \B,  \P, \S, \mathbf{\Theta} ^{-\boldsymbol{\mu}_{v, m}}) \propto\textrm{multivariate normal}(\V_m^T\mathbf{1}/(T_m + 1), \Psi_{v, m}/(T_m + 1))$.
	\item  $p(\V_m|  \B,  \P, \S, \mathbf{\Theta} ^{-\V_m}) \propto\text{matrix normal}_{T_m \times R}(\tilde{\M}_{v, m}, \mathbf{I}_{T_m}, \tilde{\Psi}_{v, m})$ 
	where 
	\begin{eqnarray*}
	\tilde{\Psi}_{v, m} &=& (\mathbf{Q}_{v, m}/\sigma_m^2 + \Psi_{v, m}^{-1})^{-1} \\
	\tilde{\M}_{v, m} &=& (\mathbf{L}_{v, m}/\sigma_m^2 + \mathbf{1}\boldsymbol{\mu}_{v, m}^{T_m}\Psi_{v, m}^{-1}) \tilde{\Psi}_{v, m}\\
	\mathbf{Q}_{v, m} &=& (\U_m^{T}\U_m)\circ(\U_m^{T}\U_m) \\
	\mathbf{L}_{v, m} &=& \sum_{i, j}b_{i, j, \cdot} \otimes (\U_{m, i, \cdot} \circ   \U_{m, j, \cdot} )
	\end{eqnarray*}
	\end{enumerate}
\item[Step 3] The sampling of $\beta$ from $\normdist{b_1}{B_1}$ where 
\begin{eqnarray*}
B_1&=& (B_0^{-1} + \sum_{m=1}^{M}\sigma^{-2}_{m} N^2 \mathbf{1}(\S = m))^{-1} \\
b_1&=& B_1 \times \Big(B_0^{-1}b_0 + \sum_{i = 1}^N\sum_{j = 1}^N \sum_{t = 1}^T b_{i, j, t} - \mu_{i, j, t} \Big).
\end{eqnarray*}
$\mathbf{1}(\S = m)$ is the number of time units allocated to state $m$ and $\mu_{i, j, t}$ is an element of $\U_{S_t}\boldsymbol{\Lambda}_t\U_{S_t}^T$.

\item[Step 4] The sampling of $\sigma^2_m$ from $\mathcal{IG}\left (\frac{c_0 +  N_m \cdot N_m \cdot T_m}{2}, \frac{d_{0} + \sum_{i = 1}^N\sum_{j = 1}^N\sum_{t=1}^T b_{i, j, t} - \beta - \mu_{i, j, t}}{2} \right)$. \end{description}

\noindent \textbf{Part 2} 
\begin{description}\footnotesize
\item[Step 5] Sample $\S$ recursively using \cite{Chib1998}'s algorithm. The joint conditional distribution of the latent states $p(S_0, \ldots, S_T | \mathbf{\Theta}, \B, \P) $ can be written as the product of $T$ numbers of independent conditional distributions: 
\begin{equation*}
 p(S_0, \ldots, S_T |\mathbf{\Theta}, \B, \P) = p(S_T| \mathbf{\Theta}, \B, \P)\ldots p(S_t|\S^{t+1},  \mathbf{\Theta}, \B, \P) \ldots p(S_0|\S^{1},  \mathbf{\Theta}, \B, \P). 
\end{equation*}

Using Bayes' Theorem, \cite{Chib1998} shows that 
\begin{eqnarray*}
p(S_t|\S^{t+1}, \mathbf{\Theta}, \B, \P) &\propto& \underbrace{p(S_t|\mathbf{\Theta}, \mathbf{B}_{1:t}, \P)}_{\text{State probabilities given all data}} \underbrace{p(S_{t+1}|S_t, \P)}_{\text{Transition probability at $t$}}.
\end{eqnarray*}
The second part on the right hand side is a one-step ahead transition probability at $t$, which can be obtained from a sampled transition matrix ($\P$). The first part on the right hand side is state probabilities given all data, which can be simulated via a forward-filtering-backward-sampling algorithm as shown in \cite{Chib1998}.

\item[Step 5-1] During the burn-in iterations, if sampled $\S$ has a state with single observation, randomly sample  $\S$ with replacement using a pre-chosen perturbation weight ($\mathbf{w}_{\mathrm{perturb}} = (w_1, \ldots, w_{M})$).  
\end{description}

\noindent \textbf{Part 3: $p(\P| \B, \S, \mathbf{\Theta})$} 
\begin{description}\footnotesize
\item[Step 6] Sample each row of $\P$ from the following Beta distribution: \\
\begin{equation*}
p_{kk} \sim \betadist{a_0 + j_{k, k} - 1}{b_{0} + j_{k, k+1}}
\end{equation*}
where $p_{kk}$ is the probability of staying when the state is $k$, and $j_{k, k}$ is the number of jumps from state $k$ to $k$, and $j_{k, k+1}$ is the number of jumps from state $k$ to $k+1$.
\end{description}

\clearpage
\section*{Model Diagnostics}

\begin{figure}
        \centering
        \includegraphics[scale = 0.34]{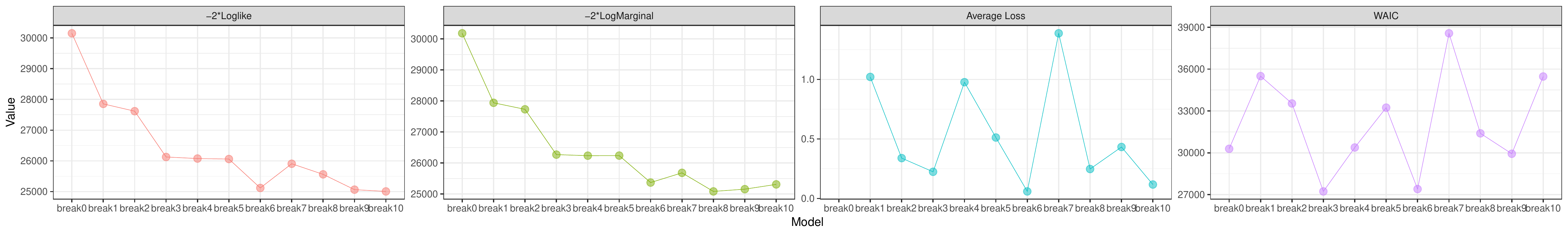}
        \caption{\emph{Model diagnostics using log likelihoods, log marginal likelihoods, average losses, and WAICs.}}
        \label{fig:diag}
    \end{figure}

Figure \ref{fig:diag} shows that average loss and WAIC prefer a HMTM with six breaks. The approximate log marginal likelihood ($\log p(\mathcal{B}|\mathcal{M}_k)$)  shows a decaying pattern as we increase the break number in general. This is a general pattern in mixture models due to the appearance of singleton states (i.e. hidden components with only one observation) in mixture models with a large number of components. We note a kink in the six break model, indicating adding one more break to the five break model improves the model fit while adding more than one break deteriorates the model fit. Note the same pattern in the log likelihoods in Figure \ref{fig:diag}.  Table \ref{marg} shows the numerical results of the approximate log marginal likelihood test. For these reasons, we determined that the six break model is most reasonable given the data. 

\begin{table}[ht]
\centering
\begin{tabular}{c r}
\toprule
Break Number & $\log p(\mathcal{B}|\mathcal{M}_k)$  \\
\midrule
 $\mathcal{M}_0$ & -15090.85 \\ 
 $\mathcal{M}_1$  & -13970.58 \\ 
 $\mathcal{M}_2$ & -13865.45 \\ 
 $\mathcal{M}_3$ & -13135.31 \\ 
 $\mathcal{M}_4$  & -13118.15 \\ 
 $\mathcal{M}_5$ & -13119.40 \\ 
 $\mathcal{M}_6$  & -12684.86 \\ 
 $\mathcal{M}_7$ & -12841.46 \\ 
 $\mathcal{M}_8$ & -12541.92 \\ 
 $\mathcal{M}_9$  & -12578.19 \\ 
 $\mathcal{M}_{10}$  & -12655.21 \\ 
 \bottomrule
\end{tabular}
   \caption{\emph{Break number detection using log marginal likelihoods.}}\label{marg}
\end{table}

\newpage
 \section*{Epoch Converter}
 Table \ref{epoch} shows the week number in our analysis and its corresponding dates.  
\begin{table}[ht]\footnotesize
\centering
\begin{tabular}{c cc}
\toprule
Week &Starting Date& Ending Date\\
\midrule
Week 1	& January 1, 2018	& January 7, 2018\\
Week 2	& January 8, 2018	& January 14, 2018\\
Week 3	& January 15, 2018	& January 21, 2018\\
Week 4	& January 22, 2018	& January 28, 2018\\
Week 5	& January 29, 2018	& February 4, 2018\\
Week 6	& February 5, 2018	& February 11, 2018\\
Week 7	& February 12, 2018	& February 18, 2018\\
Week 8	& February 19, 2018	& February 25, 2018\\
Week 9	& February 26, 2018	& March 4, 2018\\
Week 10	& March 5, 2018	& March 11, 2018\\
Week 11	& March 12, 2018	& March 18, 2018\\
Week 12	& March 19, 2018	& March 25, 2018\\
Week 13	& March 26, 2018	& April 1, 2018\\
Week 14	& April 2, 2018	& April 8, 2018\\
Week 15	& April 9, 2018	& April 15, 2018\\
Week 16	& April 16, 2018	& April 22, 2018\\
Week 17	& April 23, 2018	& April 29, 2018\\
Week 18	& April 30, 2018	& May 6, 2018\\
Week 19	& May 7, 2018	& May 13, 2018\\
Week 20	& May 14, 2018	& May 20, 2018\\
Week 21	& May 21, 2018	& May 27, 2018\\
Week 22	& May 28, 2018	& June 3, 2018\\
Week 23	& June 4, 2018	& June 10, 2018\\
Week 24	& June 11, 2018	& June 17, 2018\\
\bottomrule
\end{tabular}
   \caption{\emph{Epoch Converter}}\label{epoch}
\end{table}

\clearpage
\section*{Results}
\begin{figure}
        \centering
        \includegraphics[scale = 0.35]{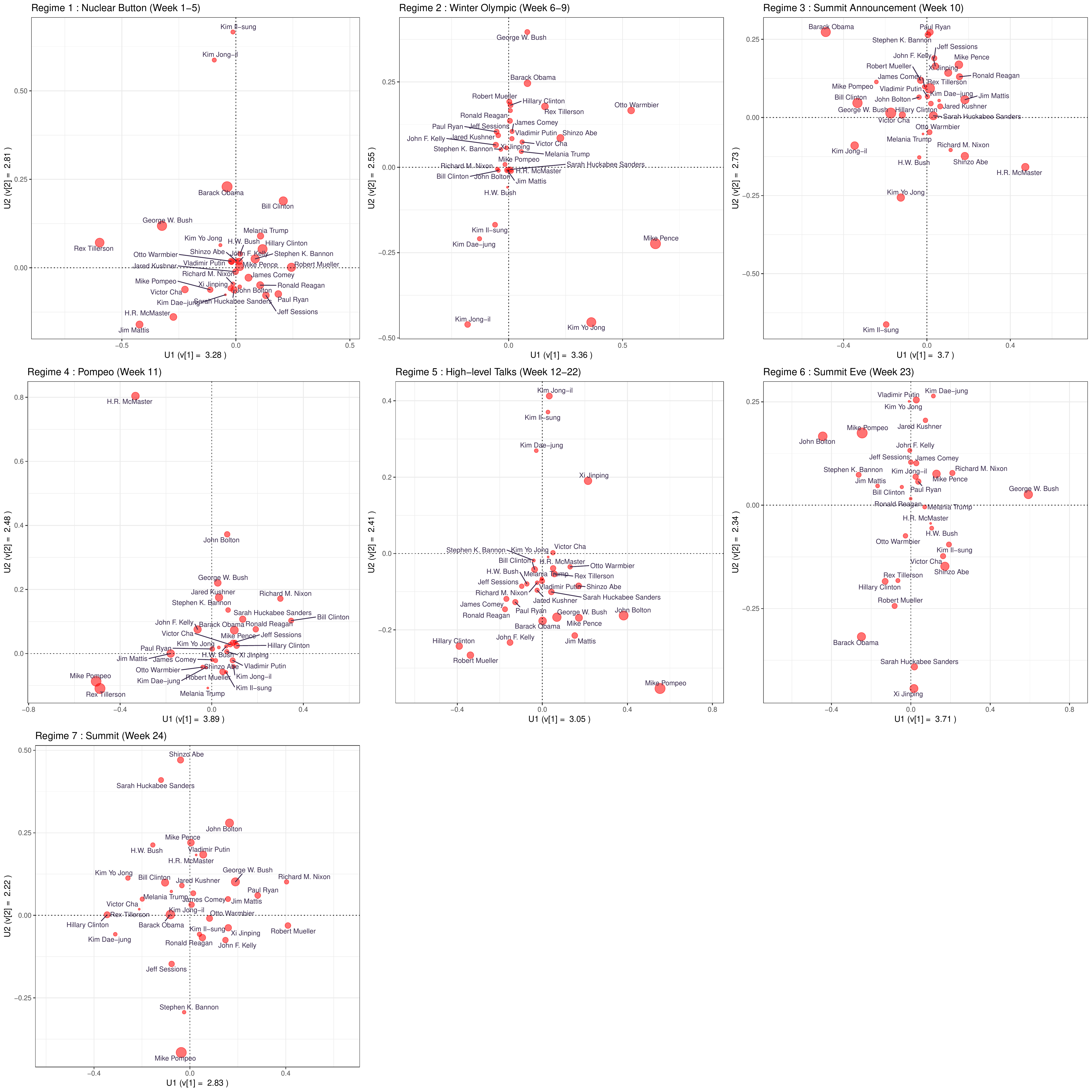}
        \caption{\emph{Changes in Latent Space of Key Figure Network}}
        \label{fig:latent}
    \end{figure}

Figure \ref{fig:latent} shows regime-specific latent node positions of key figures.

\subsection*{Discussions of Hidden Regimes}
Names of former U.S. presidents (ex. \texttt{George W.\ Bush}, \texttt{Barack Obama}, and \texttt{Bill Clinton}) very often co-occur because many news articles compared the Trump administration's approach to North Korea with the approaches of previous U.S. administrations. For example, \emph{The Washington Post} reports on March 19, 2018:  
\begin{displayquote}
\emph{``If he [President Trump] is, in fact, serious about this decision and expects to strike a deal with the North Koreans regarding their growing nuclear arsenal, he should seek the counsel of three people: former presidents Jimmy Carter and Bill Clinton, and former secretary of state Madeleine Albright. ... Though none of them were ultimately successful in persuading North Korea to commit to a meaningful long-term weapons deal, they did achieve some positive results, especially regarding the release of imprisoned Americans."} 
\end{displayquote}
\emph{The Washington Post}, ``The one thing Donald Trump should do before meeting with Kim Jong Un", March 19, 2018. Right after the summit, \emph{The Washington Post} also notes 
\begin{displayquote}
\emph{``When President Bill Clinton struck a landmark deal with North Korea in 1994, the isolated nation agreed to take a variety of specific steps that included freezing and later dismantling its nuclear program, as well as opening its facilities to international inspectors."} 
\end{displayquote}
\emph{The Washington Post}, ``Compared with Trump, previous presidents extracted more concessions from N. Korea, experts say", June 12, 2018.

\texttt{Xi Jinping} became more important during Regime 5 and Regime 6 after Kim Jong-un's surprise visit to China on March 11. Since then, President Trump had complained that China interfered in U.S.-North Korea talks.\footnote{For example, on May 18, 2018, President Trump said ``It could very well be that he[Xi Jinping]'s influencing Kim Jong-un. We'll see what happens." \emph{South China Morning Post}, ``Donald Trump says North Korea could have been `influenced' by Xi Jinping to turn on US ahead of talks with Kim Jong-un'' May 18, 2018.} Various sources reported that Trump canceled the summit to send North Korea a message that China should not meddle with the US-North Korea dialogue.

For example, according to \emph{The New York Times}, 
\begin{displayquote}
\emph{``The [summit cancellation] decision may also fan tensions between the United States and China. Mr.\ Trump has said he believes North Korea's tone changed after Mr.\ Kim met President Xi Jinping in the coastal Chinese city of Dalian in early May. Mr.\ Trump suggested that Mr.\ Xi might be using China's influence over North Korea as leverage in trade negotiations with the United States.} 
\end{displayquote}
\emph{The New York Times}, ``Trump Pulls Out of North Korea Summit Meeting With Kim Jong-un" May 24, 2018. 

And, \emph{The Wall Street Journal} reports 
\begin{displayquote}
\emph{``Mr.\ Trump considered canceling the meeting late Wednesday after a senior North Korean envoy ridiculed Vice President Mike Pence, calling him a "political dummy," and warned that Pyongyang could inflict on America an "appealing tragedy that it has never experienced nor even imagined." \ldots A U.S. intelligence official told The Wall Street Journal there was a notable shift in Mr.\ Kim's attitude immediately following his meeting with Mr.\ Xi. Intelligence officials assessed that Mr.\ Xi had some influence on that shift in an effort to place pressure on the Trump administration, the official said."}
\end{displayquote} \emph{The Wall Street Journal}, ``President Donald Trump Cancels North Korea Summit", May 24, 2018.

\clearpage
\bibliographystyle{apsr}
\bibliography{TrumpKim.bib}

\end{document}